\documentclass[11pt]{article}
\usepackage{algorithm}
\usepackage{amsfonts}
\usepackage{amsmath}
\usepackage{amssymb}
\usepackage{authblk}
\usepackage{booktabs}
\usepackage{breqn}
\usepackage[margin=1in]{geometry}
\usepackage{graphicx}
\usepackage{makecell}
\usepackage{multirow}
\usepackage{tabularx}
\usepackage{xcolor}
\usepackage[colorlinks=true,citecolor=blue,urlcolor=blue]{hyperref}
\usepackage{lineno}

\graphicspath{{./figs/}}

\title{Efficient neural topology optimization via active learning for enhancing turbulent mass transfer in fluid channels}

\author[a,b,c]{Chenhui Kou}
\author[a,d]{Yuhui Yin}
\author[c]{Min Zhu}
\author[a,*]{Shengkun Jia}
\author[a]{Yiqing Luo}
\author[a,*]{Xigang Yuan}
\author[c,*]{Lu Lu}

\affil[a]{School of Chemical Engineering and Technology, State Key Laboratory of Chemical Engineering, Tianjin University, Tianjin 300072, China}
\affil[b]{College of Chemistry \& Chemical Engineering, Yantai University, Yantai 264005, China}
\affil[c]{Department of Statistics and Data Science, Yale University, New Haven, CT 06511, USA}
\affil[d]{Department of Chemical Engineering, University College London, London WC1E 7JE, UK}
\affil[*]{Corresponding author. Email: jiask@tju.edu.cn, yuanxg@tju.edu.cn, lu.lu@yale.edu}
\date{}

\begin{document}
\maketitle
\begin{abstract}
The design of fluid channel structures of reactors or separators of chemical processes is key to enhancing the mass transfer processes inside the devices. However, the systematic design of channel topological structures is difficult for complex turbulent flows. Here, we address this challenge by developing a machine learning framework to efficiently perform topology optimization of channel structures for turbulent mass transfer. We represent a topological structure using a neural network (referred to as “neural topology”), which is optimized by employing pre-trained neural operators combined with a fine-tuning strategy with active data augmentation. The optimization is performed with two objectives: maximization of mass transfer efficiency and minimization of energy consumption, for the possible considerations of compromise between the two in real-world designs. The developed neural operator with active learning is data efficient in network training and demonstrates superior computational efficiency compared with traditional methods in obtaining optimal structures across a large design space. The optimization results are validated through experiments, proving that the optimized channel improves concentration uniformity by 37\% compared with the original channel. We also demonstrate the variation of the optimal structures with changes in inlet velocity conditions, providing a reference for designing turbulent mass-transfer devices under different operating conditions.
\end{abstract}

\paragraph{keywords:}computational fluid dynamics; mass transfer enhancement; topology optimization; neural operator; active learning; neural topology

\section{Introduction}

Turbulent mass transfer is a fundamental phenomenon and directly influences the efficiency of both separation~\cite{sanchez2014turbulent} and reaction~\cite{jin2012mass,ramirez2008mass}, the two major operations in chemical and many other process industries~\cite{sundararajan2009turbulent}. Enhancing mass transfer is an effective way of process enhancement aiming at inherently and significantly increasing the efficiency of the processes~\cite{zhang2019optimization}. Experimental studies showed that mass transfer is strongly affected by the inside geometric structure of the channel of the fluid flow, as, in addition to molecular diffusion, its convection and fluctuation that drive the movement of species within the fluid mixture are determined mainly by the structure~\cite{kumar2014experimental,tao2017experimental,hassan2020mass,lu2024flow}.

Efforts have been made by altering fluid channels’ structures in process equipment to devise structures that intensify the convection and fluctuation-driven mass transfer in turbulent mass transfer processes. These include addition of baffles~\cite{chen2008field}, porous media~\cite{bidi2010numerical}, and installing elements to improve flow patterns or to induce disturbances~\cite{zheng2016flow,kou2022performance,nishimura2000influence}. However, in most of the previous research, the designs of improved structures were based on designers’ imaginations or inspirations combined with experimental and/or numerical validations, and, therefore, the best structure giving the most effective turbulent mass transfer cannot be guaranteed.

Topology optimization (TO) is a systematic way of finding the best structure for given objectives, including, for example, mass transfer flux maximization and flow resistance minimization~\cite{bendsoe2003topology,allaire2002topology}. Since mass transfer typically occurs in turbulent flows within chemical equipment and optimal topology is easily affected by velocity boundary conditions, optimization of the structure of the fluid channel by TO to improve turbulent mass transfer efficiency is a challenge~\cite{dilgen2018topology,wang2020topology}.
In recent years, researchers have applied TO to mass transfer processes such as reactors~\cite{jia2021multi,cao2020optimization} and micro-mixing~\cite{xie2017numerical,ansari2010mixing}. These studies typically formulated the optimization problem employing a Lagrangian augmentation equation based on a mechanism model, objective functions, and process constraints, and then solved the problem using algorithms such as the method of moving asymptotes (MMA) or genetic algorithms. Since the turbulent mass transfer process is complex, these methods, which require repeatedly solving a mechanism model and Lagrange multiplier equations, demand significant computational resources~\cite{mikkelsen2004topology,kou2024physics}. Moreover, these methods are limited to optimizations under a single boundary condition and cannot take into account varying boundary conditions, such as inlet velocities, which have an impact on the optimal topological structure. 

The recent development of scientific machine learning methods~\cite{karniadakis2021physics}, such as physics-informed neural networks~\cite{raissi2017physics,raissi2020hidden,fan2024deep,daneker2024transfer,wu2024identifying} and neural operators~\cite{lu2021learning,jin2022mionet,lu2022comprehensive,jiao2021one,jiao2024solving,zhu2023reliable}, has provided effective approaches for physical fields prediction under different operation conditions. Researchers have applied these models to fluid optimization, including streamline design of flow reactor~\cite{savage2024machine}, airfoil~\cite{yang2024buffet,yang2024transferable,yang2024rapid}, artificial catheters~\cite{zhou2024}, solid location optimization in nanoscale heat transfer processes~\cite{lu2022multifidelity}, and solid shape optimization within laminar flow channels~\cite{lu2021physics,hao2022bi}. However, these studies primarily focus on optimizing specific geometric parameters within fluid channels, and large training datasets are typically required for flow channel structure optimization. To find optimal topological structures within high-dimensional function spaces (e.g., in the case of fluid flow channel structure optimization to enhance turbulent mass transfer), the volume, position, and shape of solid structures within channels need all be handled as variables, and the impact of varying velocity boundaries on the optimal fluid channel design should also be considered. However, such a method is not available yet in the open literature.

In this study, we represent a topological structure using a neural network (``neural topology''), which is integrated with pre-trained neural operators to realize topology optimization under different inlet velocities for turbulent mass transfer. The recently proposed Fourier-enhanced DeepONet (Fourier-DeepONet)~\cite{zhu2023fourier,jiang2024fourier,lee2024efficient}, known for its robust generalization ability and training efficiency, is used to construct neural operators, which can efficiently predict concentration and pressure distributions under different topological structures and inlet velocities. Process indicators are defined based on the predicted concentration and pressure distributions, serving as objectives for the gradient-based optimization of a neural topology under varying inlet velocities using a neural network. Through two rounds of active data augmentation based on optimization results, both neural operators and the optimization results are improved, enabling data-efficient optimization within a larger and complex design space. Compared with the traditional mechanism model-based optimization approach, the proposed method in this study demonstrates superior computational efficiency under varying boundary conditions. We also design turbulent mass transfer experiments of the optimized channel to validate the effectiveness of the proposed TO method. The results under different inlet velocities are quantitatively compared and analyzed to guide the channel design of the actual turbulent mass transfer process.

\section{Results}

\begin{figure}[htbp]
    \centering
    \includegraphics[width=1\linewidth]{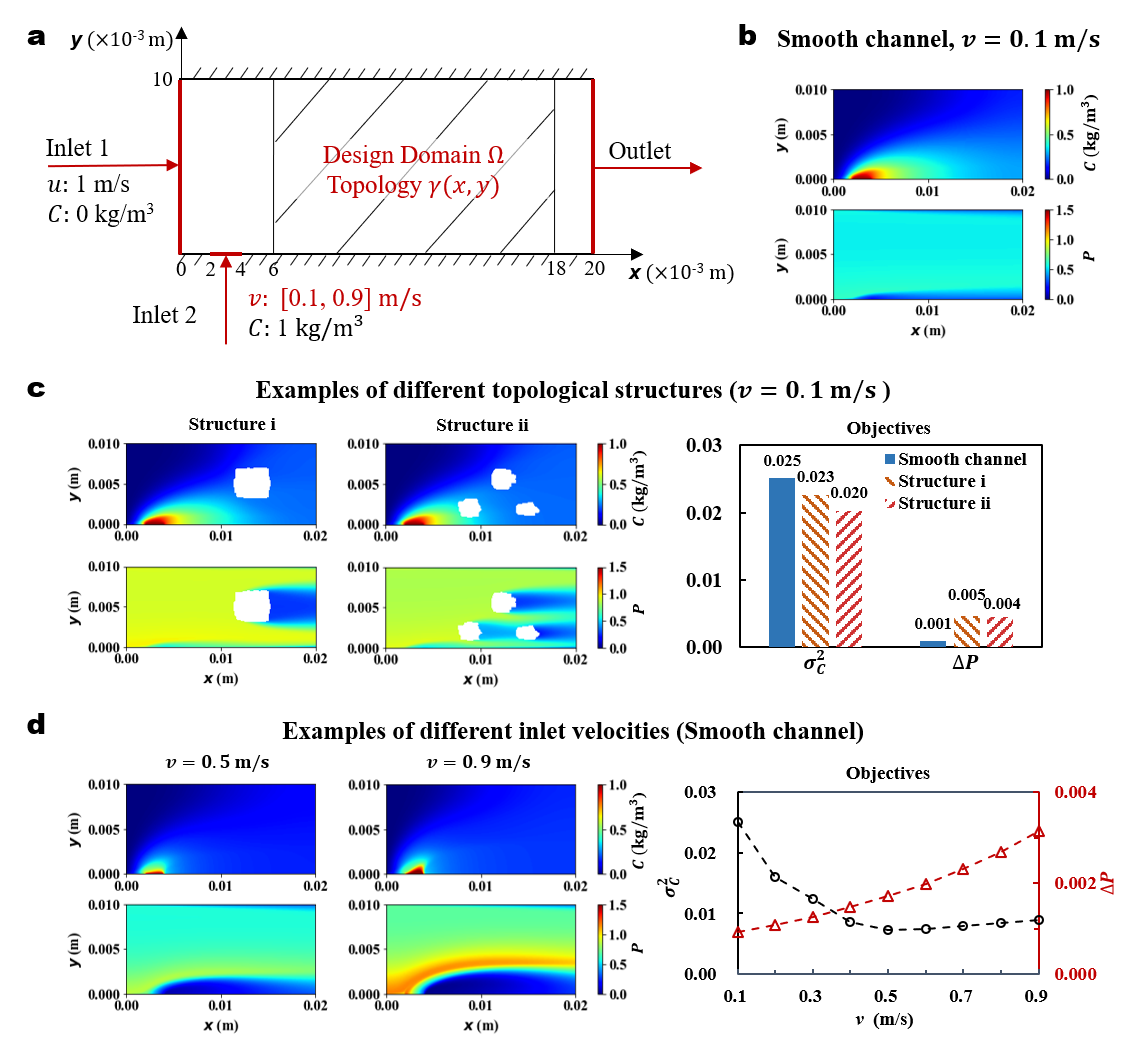}
    \caption{\textbf{Effects of topology and inlet velocity on turbulent mass transfer.} \textbf{a}, The computational domain of the turbulent mass transfer process for a water (solvent)--methylene blue (solute) system. \textbf{b}, The concentration and pressure distributions of the smooth channel under \(v=0.1\,\mathrm{m/s}\). \textbf{c}, The concentration and pressure distributions of the topological structure i and structure ii, and comparison of objective values between the smooth channel and example structures. \textbf{d}, Concentration and pressure distributions under \(v=0.5\,\mathrm{m/s}\) and \(v=0.9\,\mathrm{m/s}\), and objective values for different inlet velocity.}
    \label{fig:main1}
\end{figure}

This study considers the turbulent mass transfer process for a water (solvent)--methylene blue (solute) system in a rectangular channel~\cite{kou2024physics} (Fig.~\ref{fig:main1}a). The topology optimization is performed in a rectangular domain 
\(\Omega=\left\{(x,y)\text{ }\middle|\text{ }0.006<x<0.018,\text{ } 0<y\allowbreak <0.01\right\}\) in the channel. The effect of inlet velocity $v$ in the range [0.1, 0.9] on the process efficiency and the optimal topological structure is also investigated. We suppose that the space in the channel is full of solids with pores and the channel’s structure is altered by changing its porosity distribution $\gamma\left(x,y\right)$. To ensure the effectiveness of the generated topological structure, the following solid volume constraint is imposed:
\begin{equation}
\frac{1}{V_\Omega} \iint_\Omega {(1-\gamma)}dxdy \geq \theta, \label{eq:10}
\end{equation}
where \( \theta = 10\% \) is the minimum solid volume fraction in the design domain, and \( V_\Omega \) is the total volume of the design domain.

Most topology optimization research focuses on enhancing mass transfer as the primary objective. However, channels with better mass transfer may require higher external energy consumption, which could lead to a decline in the overall performance of the process. Therefore, our optimization process employs multi-objective optimization, with the process efficiency indicated by two criteria:
\begin{equation}
\sigma_C^2 = \frac{1}{V_\Omega} \iint_\Omega {\left(C(x,y)-\bar{C}\right)}^2 dxdy, \label{eq:7}
\end{equation}
\begin{equation}
\Delta P = \int_{\text{inlet 1}} P \, dy+\int_{\text{inlet 2}} P \, dx -\int_{\text{outlet}} P \, dy,
\label{eq:8}
\end{equation}
where \( \sigma_C^2 \) represents the concentration ($C$) variance of the system and reflects the uniformity of the concentration of the methylene blue, indicating the effectiveness of mass transfer; and \( \Delta P \) denotes the pressure ($P$) drop from the inlet to the outlet of the system, representing the external energy consumption for the process. \(\bar{C}\) represents the average concentration of methylene blue in the fluid channel.

We aim to identify the optimal topological structure of the channel that balances mass transfer efficiency and external energy consumption for the process with different velocity boundaries. The total objective function \( Obj \) of the optimization is defined by the weighted sum of \( \sigma_C^2 \) and \( \Delta P \):
\begin{equation}
Obj=\sigma_C^2+\omega\cdot\Delta P, \label{eq:9}
\end{equation}
where \( \omega \) represents the weight. Such a best channel structure can be either mass transfer effectiveness dominant or energy efficiency dominant, depending on the weight value given by the decision-making designers.
 
\subsection{Effects of topology and inlet velocity on turbulent mass transfer}

We first investigate the effects of topology and inlet velocity on turbulent mass transfer. The concentration and pressure distributions for the smooth channel and channels with randomly generated topological structures (Figs.~\ref{fig:main1}b and c) are numerically simulated based on the mechanism model presented in Section~\ref{sec:mechanism-model}. Comparisons between the results of the three fluid channels (Fig.~\ref{fig:main1}c) show that modifying the flow channel by adding solid blocks generally results in a more uniform concentration distribution (smaller \( \sigma_C^2 \)) but with the cost of increase of the system pressure drop (larger \( \Delta P \)) compared to the smooth channel. The Structure ii achieves smaller values of both \( \Delta P \) and \( \sigma_C^2 \) compared to that of Structure i, and therefore Structure ii is more favorable for the mass transfer process. This demonstrates that different channel structures behave differently in terms of the two criteria, and optimization is necessary to find the best topological structure of the channel that minimizes the overall objective function.

The inlet velocity of the system is also a crucial factor influencing mass transfer efficiency. The system pressure drop \( \Delta P \) increases with the increase of inlet velocity (Fig.~\ref{fig:main1}d), whereas the \( \sigma_C^2 \) decreases first and then increases with the inlet velocity. The concentration distribution is the most uniform at \(v=0.5\,\mathrm{m/s}\). This suggests that increasing the inlet velocity improves both the convective and fluctuating mass transfer in the turbulent flow system. At the same time, the flow rate of the solute entering the system also increases with the rise of inlet velocity \(v\). Therefore, the high-concentration region becomes larger at \(v=0.9\,\mathrm{m/s}\)  compared to \(v=0.1\,\mathrm{m/s}\) , leading to an increase in the concentration variance.

\subsection{Optimizing neural topology under different inlet velocities}
\label{optimization method}

\begin{figure}[htbp]
    \centering
    \includegraphics[width=1\linewidth]{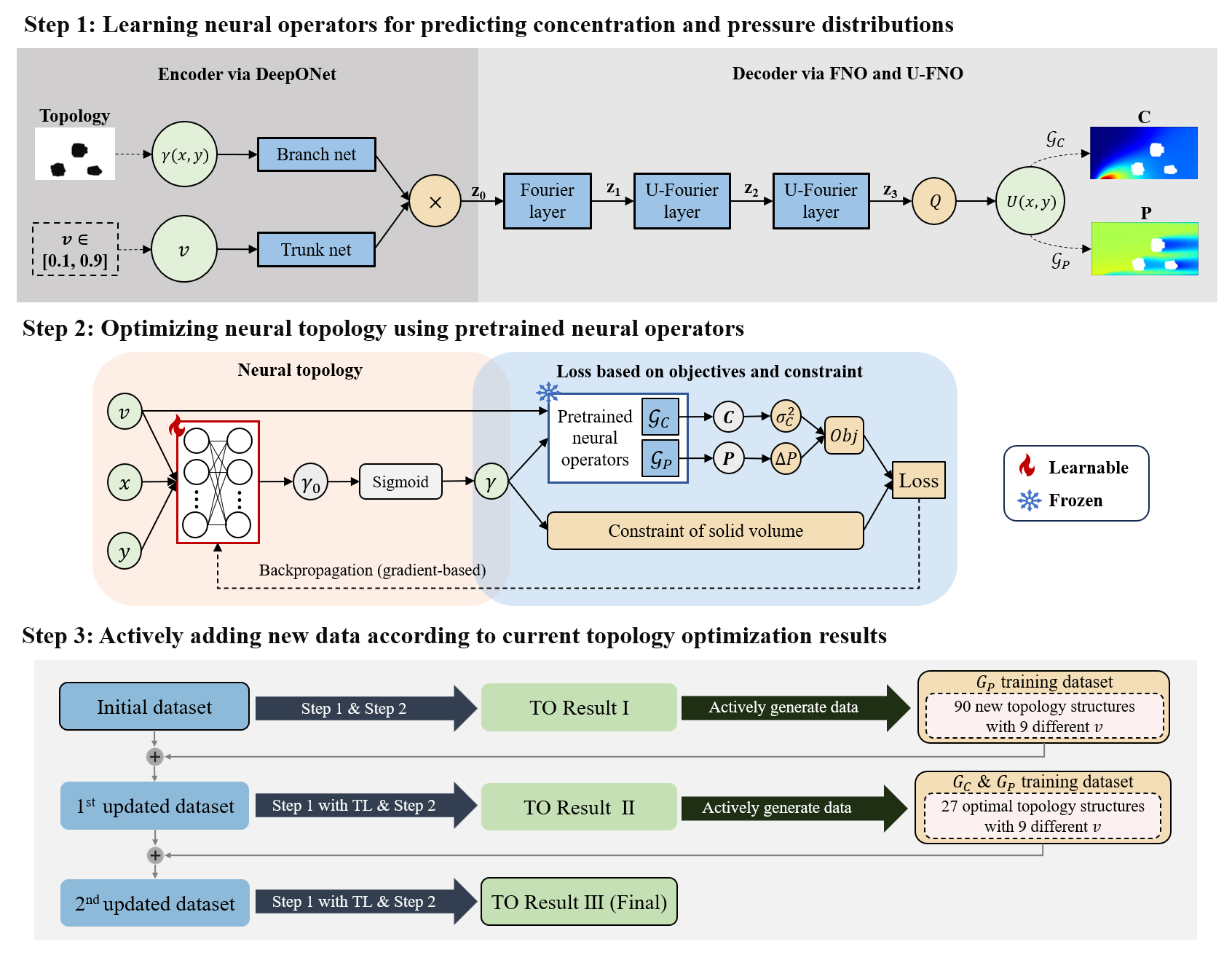}
    \caption{\textbf{Algorithm flowchart.} \textbf{Step 1}, Learning neural operators for predicting concentration and pressure distributions. The neural operator maps the topological structure \( \gamma(x,y) \) and inlet velocity $v$ to physical fields \( U(x,y) \). The output variable \( U \) represents \( C \) or \( P \). \textbf{Step 2}, Optimizing a neural topology (a neural network representing a topological structure)  under different inlet velocities using pre-trained neural operators. \textbf{Step 3}, Actively adding new data according to current topology optimization results to fine-tune the optimization results.}
    \label{fig:main2}
\end{figure}

Since the governing equations based on the mechanism model for the turbulent mass transfer process are complex, traditional methods cannot efficiently optimize the channel structure, especially for the investigations on various inlet velocities. In the present study, we develop a computational framework to explore the optimal topology under different inlet velocities by integrating pre-trained neural operators with a neural topology (Fig.~\ref{fig:main2}). To address the challenge of the large search space of the possible topological structures, we propose an active data-augmentation method to enhance the effectiveness of the neural operator-based optimization algorithm. The method can be described by the following three steps.

\paragraph{Step 1.}
Two independent neural networks $\mathcal{G}_C$ and $\mathcal{G}_P$ are trained for rapid prediction of the concentration distribution $C(x,y)$ and pressure distribution $P(x,y)$ under different topological structures $\gamma\left(x,y\right)$ and inlet velocity $v$, i.e., $$C\left(x,y\right)=\mathcal{G}_C(\gamma\left(x,y\right),v) \quad \text{and} \quad P\left(x,y\right)=\mathcal{G}_P(\gamma\left(x,y\right),v).$$
In this study, the neural operator method of Fourier-DeepONet~\cite{zhu2023fourier} is adopted to construct the neural networks $\mathcal{G}_C$ and $\mathcal{G}_P$, which demonstrated high training efficiency and strong generalization capability. In Fourier-DeepONet, the encoding process by DeepONet consists of two fully-connected neural networks: the branch net used for encoding the input function \( \gamma(x,y) \) and the trunk net used for encoding the input variable \( v \). The output of DeepONet goes through the decoding process of a Fourier layer, 2 U-Fourier layers, and a projection layer \( Q \), to get the output function of the Fourier-DeepONet framework. The details of the training data generation and neural networks are given in Sections~\ref{sec:data-generation} and \ref{sec:neural operator}. 

\paragraph{Step 2.}
To simultaneously perform TO under different velocity boundary conditions, we constructed a neural topology capable of outputting the topological structure for any given inlet velocity \( v \). Within the TO framework, the pre-trained neural operators \( \mathcal{G}_C \) and \( \mathcal{G}_P \) are integrated to efficiently compute the corresponding objective values for any given topological structure \( \gamma(x,y) \) and inlet velocity \( v \). The objective function (Eq.~\eqref{eq:9}) together with the inequality constraint of solid volume (Eq.~\eqref{eq:10}) forms the total loss \( \mathcal{L} \) of the TO framework. Then, a gradient-based algorithm is used to optimize the parameters of the neural topology. As such, the trained neural topology can output optimized topological structures \( \gamma(x,y) \) with minimum \( \mathcal{L} \) for any given inlet velocity. The algorithm is detailed in Section~\ref{sec:optimization algorithm}.

\paragraph{Step 3.}
Due to the large search space of the topological structures, the neural operator trained on the randomly generated initial training dataset may not achieve sufficient prediction accuracy for all possible topological structures. This may result in discrepancies between the optimal topological structures obtained from the neural operator-based TO algorithm and those derived from the mechanism model-based TO algorithm. To address this issue, we develop an active learning method that performs two rounds of data-augmentation process according to the current TO results. The details of the active data-augmentation methods are discussed in Section~\ref{sec:active learning}. Using such a data-augmentation method, both neural operators and the TO framework were fine-tuned to obtain the final TO results, which are then validated using the mechanism model.

\subsection{Optimal topology for different objectives}

\begin{figure}[htbp]
    \centering
    \includegraphics[width=1\linewidth]{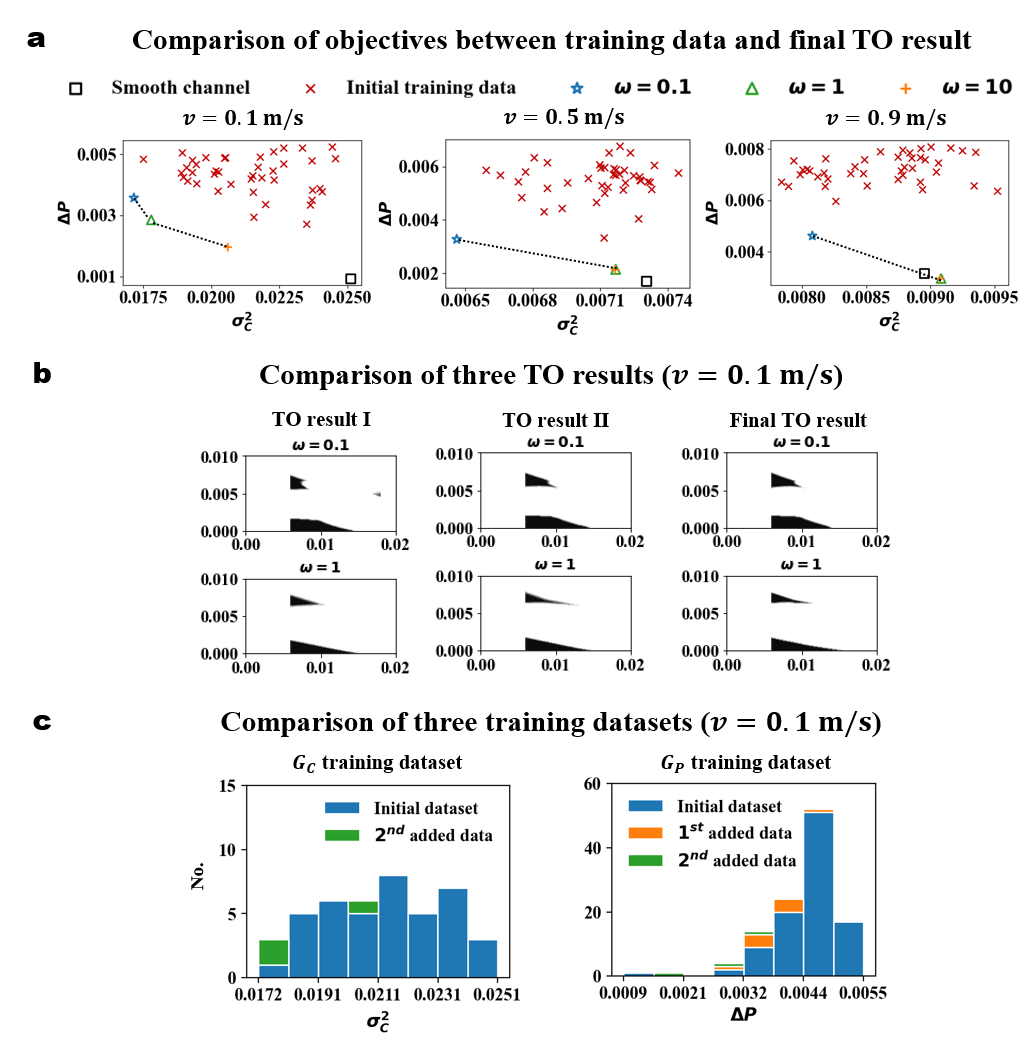}
    \caption{\textbf{Optimal topology and corresponding objectives.} \textbf{a}, Comparison of objective values between training dataset and TO results under \(v=0.1 \,\mathrm{m/s}\) , \(0.5 \,\mathrm{m/s}\) , and \(0.9 \,\mathrm{m/s}\) . The dashed lines represent the possible Pareto front. \textbf{b}, Comparison of optimal topological structures between three TO results for $v=0.1\,\mathrm{m/s}$ and \(\omega=0.1\) or 1. \textbf{c}, Objective distributions of \( \mathcal{G}_C \) training datasets and \( \mathcal{G}_P \) training datasets with $v=0.1\,\mathrm{m/s}$ during active learning.}
    \label{fig:main4}
\end{figure}

After introducing the methods, we present the TO results under different inlet velocities. We first compare the objective values between the cases in the training dataset and the TO results in Fig.~\ref{fig:main4}a. Specifically, we show the optimization results under three weights: $\omega = 0.1$, 1, and 10. These weights can be interpreted as optimization processes dominated by enhancing mass transfer performance, balanced optimization of both objective functions, and optimization driven by minimizing external energy consumption, respectively. The dashed lines in Fig.~\ref{fig:main4}a represent a possible Pareto front, which is obtained by linearly interpolating the optimization results under three different weights. When \(\omega=0.1\), the optimization algorithm achieves a channel structure with the smallest \(\sigma_C^2\) compared to the examples in the training set, especially for lower inlet velocity. In addition, the system pressure drop increases significantly as the inlet velocity increases, which is in agreement with the simulation results in Fig.~\ref{fig:main1}d. As a result, reducing \(\sigma_C^2\) no longer dominates absolutely when \(v=0.9\,\mathrm{m/s}\) , and the optimal topological structure is a trade-off between the two objectives.

As \(\omega\) and \(v\) increase, the optimization process gradually shifts from being predominantly driven by reducing \(\sigma_C^2\) to being predominantly driven by reducing \(\Delta P\). Consequently, the optimal points move towards the lower right corner in Fig.~\ref{fig:main4}a. When \(v=0.9\,\mathrm{m/s}\) , the \(\Delta P\) corresponding to the optimal topology with \(\omega=1\) and 10 are close to \(\Delta P\) of the smooth channel. These results indicate the effectiveness of our proposed TO method, which can output a topology with smaller \(\Delta P\) when the TO objective is dominated by the system pressure drop.

Next, we compare the optimized topological structures for $v = 0.1\,\mathrm{m/s}$ before and after data augmentation (Fig.~\ref{fig:main4}b). When $\omega = 1$, the optimal topological structures generated by the TO algorithm before and after data augmentation are quite similar (Fig.~\ref{fig:main4}b, second row). However, in the case of $\omega = 0.1$, the optimal topological structure obtained after data augmentation shows a noticeable difference (Fig.~\ref{fig:main4}b, first row). More optimization results (Fig.~\ref{fig:A3}) show that for most of the examples the proposed method is successful in getting a better result after one active data augmentation, and the second data augmentation can further fine-tune the optimization results.

As discussed above, we obtained good TO results even in the first round without data augmentation. This good performance is attributed to the good prediction accuracy and generalization ability of the neural operators trained only with the initial small dataset. The neural operators can accurately predict process objectives under different inlet velocities or topological structures. In our training dataset (Appendix~\ref{ap:Initial dataset}), we use $v\in\left\{0.1,0.2,\dots0.9\right\}\,\mathrm{m/s}$. To quantify the generalization ability, we design three datasets for testing, including a standard testing dataset with $v\in\left\{0.1,0.2,\dots0.9\right\} \,\mathrm{m/s}$, an interpolation dataset with $v\in [0.1, 0.9] \,\mathrm{m/s}$, and a difficult extrapolation dataset with $v \in\left\{0.05,0.95\right\} \,\mathrm{m/s}$. The prediction errors of the neural operators for three test datasets (Table~\ref{tab:A1}) are less than 1\%, 2\% and 5\%, respectively. Hence, the trained \( \mathcal{G}_P \) and \( \mathcal{G}_C \) neural operators can not only predict the new topological structure, but also has good prediction accuracy for the cases whose inlet velocities are outside the training set. Moreover, although the neural operators are not directly trained with the objective values, the objective values calculated using the predicted physical fields are in good agreement with the results from the mechanism model. We present more details on neural operator validation in Appendix~\ref{ap:NO validation}.

\begin{table}[htbp]
\caption{\textbf{Generalization test of neural operators $\mathcal{G}_C$ and $\mathcal{G}_P$ trained by the initial training dataset.} The numbers are relative errors for physical fields and objectives. Three datasets are used for testing: in the ``Test'' dataset, $v\in\left\{0.1,0.2,\dots0.9\right\} \,\mathrm{m/s}$, which is used in the training dataset; in the ``Interpolation'' dataset, $v\notin \left\{0.1,0.2,\dots0.9\right\} \,\mathrm{m/s}$; and in the ``Extrapolation'' dataset, $v \notin[0.1,0.9] \,\mathrm{m/s}$.}
\centering
\begin{tabular}{cccccc}
\toprule
Neural operator &  & Test  & Interpolation & Extrapolation \\
\midrule
\multirow{2}{*}{$\mathcal{G}_C$} 
& $C$ & 0.395\% & 1.688\% & 3.972\%  \\
& $\sigma_C^2$ & 0.293\% & 1.314\% & 4.911\%  \\
\midrule
\multirow{2}{*}{$\mathcal{G}_P$} 
& $P$ & 0.819\% & 1.851\% & 2.811\%  \\
& $\Delta P$ & 0.124\% & 1.971\% & 1.434\%  \\
\bottomrule
\end{tabular}
\label{tab:A1}
\end{table}

Moreover, we illustrate how data augmentation, particularly for the $\mathcal{G}_P$ training dataset, allows the TO framework to achieve better topology. The pressure difference \(\Delta P\) for the data points in the initial \(\mathcal{G}_P\) training dataset mostly ranges between 0.004 and 0.005 (Fig.~\ref{fig:main4}a), while in the TO results, it is approximately 0.003 for \(\omega = 0.1\), 0.002 for \(\omega = 1\), and 0.001 for \(\omega = 10\).
Excluding the smooth channel, the minimum value of $\Delta P$ in the initial training dataset is around 0.003. Therefore, the training dataset of the $\mathcal{G}_P$ has fewer data points near the optimal point, leading to lower prediction accuracy for the optimized structure. In contrast, for the $\mathcal{G}_C$ neural operator (Fig.~\ref{fig:main4}c), where the distribution of $\sigma_C^2$ in the training dataset is more uniform, the prediction of neural operator for optimized structure is more accurate. After adding more data points to the training set based on the TO results, the number of data points with smaller objective values increases (Fig.~\ref{fig:main4}c), which improves the prediction accuracy of neural operators for optimized structure.

\subsection{Computational efficiency}

We make a comparison of the computational cost between our proposed method and the traditional method. Although our method requires solving the computational fluid dynamics (CFD) and computational mass transfer (CMT) model equations repeatedly for various conditions to establish the training dataset, each simulation is independent and can be performed in parallel. Based on the trained neural operator, the optimal structure of 9 inlet velocities can be obtained by the optimization algorithm. When the $\omega$ in the objective function changes, it is only necessary to repeat Step 3 to get a new optimization result. In contrast, structure optimization based on mechanism models requires solving multiple CMT and Lagrange equations iteratively. For example, Jia et al.~\cite{jia2021multi} employed the MMA optimization algorithm and used 31 iterations, in each of which the mechanism models equations must be solved, to obtain the optimal channel structure with fixed values of $v$ and $\omega$.

For numerical solutions with the mechanism model, we utilizes parallel computing with 48 threads on two Intel Xeon E5-2687W v4 CPUs. The neural network training is performed using an NVIDIA GeForce RTX 3090 Ti GPU. The computational cost is summarized in Table~\ref{tab:9}, which shows that the proposed method has an obvious computational advantage in optimizing multiple structures under different inlet velocities. Furthermore, as we showed in Table~\ref{tab:A1}, the trained neural operator has sufficient prediction accuracy for the physical fields and objectives corresponding to any $v\in[0.1,0.9] \,\mathrm{m/s}$. Hence, if we aim to perform TO for more inlet velocities, the speedup will be even more significant.

\begin{table}[htbp]
\centering
\caption{\textbf{Comparison of computational cost between our method and the traditional mechanism model-based method.} The computation time is determined based on the TO results for 27 different conditions ($v\in\left\{0.1,0.2,\dots0.9\right\}\,\mathrm{m/s}$ and $\omega\in\left\{0.1,1,10\right\}$).}
\label{tab:9}
\begin{tabular}{cl}
\toprule
\multirow{7}{*}{\makecell{Our method \\(neural operator + \\gradient-based algorithm)}} 
& \makecell[l]{Total time: 43.5 h} \\  
& \makecell[l]{\textbullet\ Initial data generation: 6.7 h for solving CMT equations 360 \\\hspace{0.8em}times, and 3 h for solving CFD equations 540 times.} \\
& \makecell[l]{\textbullet\ Step 1: 2.8 h for \( \mathcal{G}_P \), and 2.5 h for \( \mathcal{G}_C \).} \\
& \makecell[l]{\textbullet\ Step 2: 3 h for a fixed $\omega$.} \\
& \makecell[l]{\textbullet\ Step 3: 10 h form TO result I to TO result II, and 9.5 h from\\\hspace{0.5em} TO result II to final TO result.} \\
\midrule
\multirow{4}{*}{\makecell{Traditional method \\(mechanism model + MMA)}} 
& Total time: 1080 h \\
& \makecell[l]{\textbullet\ For fixed $v$ and $\omega$, about 40 h for solving CMT and Lagrange   \\\hspace{0.5em} equations 31 times~\cite{jia2021multi}.}\\
& \makecell[l]{\textbullet\ Repeating the above process for 27 conditions of different $v$ or $\omega$}. \\
\bottomrule
\end{tabular}
\end{table}

\subsection{Experimental validation}
\label{Experiment setup}

\begin{figure}[htbp]
    \centering
    \includegraphics[width=1\linewidth]{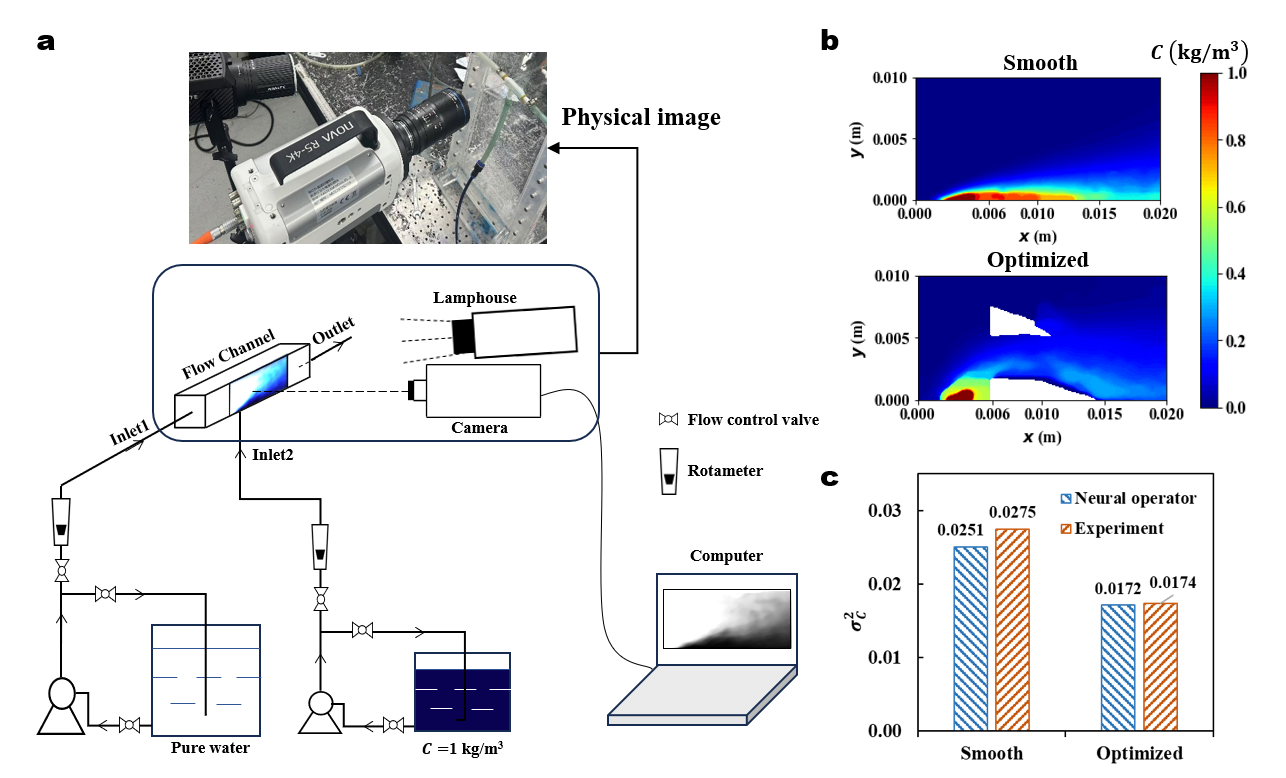}
    \caption{\textbf{Turbulent mass transfer experiment.} \textbf{a}, Schematic diagram of the experiment setup for turbulent mass transfer process. \textbf{b}, Concentration distributions of the smooth channel and optimized channel. \textbf{c}, Comparison of $\sigma_C^2$ between neural operator predictions and experiments.}
    \label{fig:main5}
\end{figure}

To validate our ML results, we conduct turbulent mass transfer experiments of the smooth and optimized flow channels under \( v = 0.1 \) m/s and \( \omega = 0.1 \). Fig.~\ref{fig:main5}a shows the diagram of the experiment setup. The simulated 2D channel was expanded into a 3D channel with a square cross-section. Three sides of the 3D channel were printed using white material, while the remaining side was fitted with a transparent window to capture the mass transfer behavior of water-methylene blue within the channel. The dimensions of the transparent window match those of the computational domain in Fig.~\ref{fig:main1}a. A centrifugal pump was used to achieve continuous inlet feed, with the inlet velocity controlled by a rotor flowmeter and a bypass control valve. During the experiment, a high-speed camera continuously captured the fluid mixing conditions within the channel, and the images were processed into grayscale on a computer. By averaging the grayscale values across all frames in the animation, the time-averaged grayscale images were obtained. The time-averaged concentration distribution (Fig.~\ref{fig:main5}b) was then obtained using the time-averaged grayscale image and the calibration curve. The original experimental data and calibration curves are presented in Appendix~\ref{ap:experiment}.

In order to quantitatively verify the effectiveness of the optimized channel in enhancing mass transfer, the variance of concentration distributions  ($\sigma_C^2$) for both the smooth and the optimized fluid channels are compared in Fig.~\ref{fig:main5}c. Experimental results indicate that the mass transfer performance of the optimized fluid channel is significantly improved compared to the smooth case, with the objective function value decreasing by approximately $37\%$, which aligns well with the simulation results. 

\subsection{Influence of inlet velocity on optimal topological structures and objectives}

\begin{figure}[htbp]
    \centering
    \includegraphics[width=0.8\linewidth]{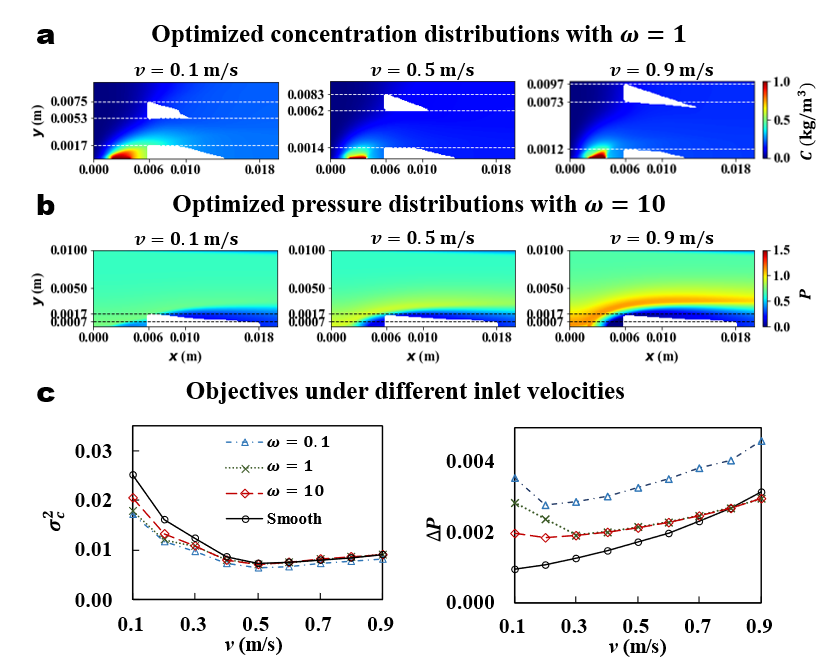}
    \caption{\textbf{Influence of inlet velocity on TO results.} \textbf{a}, Optimized concentration distributions with \(\omega=0.1\) under different inlet velocities. \textbf{b}, Optimized pressure distributions with \(\omega=10\) under different inlet velocities. \textbf{c}, Objective values of $\sigma_C^2$ and $\Delta P$ for different inlet velocities.}
    \label{fig:main6}
\end{figure}

According to the TO results in Fig.~\ref{fig:main4}a, it is evident that the optimal topological structure varies with the inlet velocity. Here, we present more analysis in two cases: $\omega=0.1$ (mass transfer dominated; Fig.~\ref{fig:main6}a) and $\omega=10$ (energy consumption dominated; Fig.~\ref{fig:main6}b).

When \(\omega=0.1\), the primary objective of the optimization process is to minimize $\sigma_C^2$, and we show the optimized concentration distributions for three different inlet velocities in Fig.~\ref{fig:main6}a. We find that the vertical location of the solid baffle increases with $v$, which is consistent with the position where the concentration drops to 0 in Figs.~\ref{fig:main1}b and d. This indicates that adding solid baffles enhances both the turbulent diffusion coefficient and the convective diffusion rate, which promotes the diffusion of solute toward regions of lower concentration, resulting in higher mass transfer rates and a more uniform concentration distribution. Another solid baffle is located at the lower wall, where the concentration boundary layer is more pronounced. As the thickness of the concentration boundary layer at this location decreases with the increase of $v$, the volume of the solid in this region also decreases accordingly.

For \(\omega=10\), the primary objective of the TO is to reduce the $\Delta P$. The optimal topological structures (Fig.~\ref{fig:main6}b) indicate that when the solid baffle is located at the bottom of the design domain $\Omega$, the pressure distribution of the process is similar to that of the smooth channel in Figs.~\ref{fig:main1}b and d. This can be explained as that a relatively thick pressure boundary layer forms near the bottom wall due to the presence of inlet 2. When the solid is positioned near the bottom of the domain, its impact on the system’s pressure drop is relatively small. Since the position of the pressure boundary layer remains unchanged with variations in $v$, when the optimization process is dominated by minimizing $\Delta P$, the optimal topological structure is independent of $v$.

We compare the objective values before and after the TO in Fig.~\ref{fig:main6}c. When the value of $\omega$ is relatively small, the process performance, controlled by reducing the concentration distribution variance, is significantly improved by the TO compared to the smooth channel. Moreover, for smaller $v$, the increase in system efficiency by the TO is more pronounced. The results for a larger value of the \(\omega\) show that the external energy consumption of the smooth channel is lower than that of the optimized channel for inlet velocities \(v < 0.8\,\mathrm{m/s}\). However, for higher inlet velocities, the pressure distribution by the TO outperforms that of the smooth channel.

\section{Discussion} 

In this study, we develop a computational framework to achieve efficient topology optimization under different velocity boundaries by integrating pre-trained neural operators with a neural topology and gradient-based optimization. By incorporating an active data augmentation approach, our framework is both data-efficient and computationally efficient in identifying the optimal solution within a large design space. Our results show that no matter whether the topology optimization is dominated by the mass transfer efficiency or the system pressure drop, the optimized topological structure is always better than the baseline smooth channel and all the randomly generated structures in the training dataset. The TO results are used to guide the design of the experiments, and we have a close agreement between the ML predictions and experimental measurements. The optimized channel shows approximately a 37\% improvement in mass transfer efficiency compared to the smooth channel. Our developed framework can be applied to the design of other turbulent mass transfer enhancement techniques, such as incorporating reactive particles and altering the fluid inlet angle.

\section{Methods}
\label{sec:methods}

\subsection{Mechanism model}
\label{sec:mechanism-model}

The turbulent mass transfer process (Fig.~\ref{fig:main1}a) can be described by the Reynolds-averaged Navier-Stokes (RANS) equations~\cite{jia2022renormalization,batchelor1967introduction} with $i$ denoting $x$ and $y$ coordinates:
\begin{equation}
\frac{\partial\rho u_i}{\partial \mathbf{x}_i}=0, \label{eq:1}
\end{equation}
\begin{equation}
u_i\frac{\partial\rho u_j}{\partial \mathbf{x}_i}=-\frac{\partial p}{\partial \mathbf{x}_j}+\frac{\partial}{\partial \mathbf{x}_i}\left(\left(\mu_m+\mu_t\right)\frac{\partial u_j}{\partial \mathbf{x}_i}\right)+f_i, \label{eq:2}
\end{equation}
\begin{equation}
u_i\frac{\partial\rho C}{\partial \mathbf{x}_i}=-\frac{\partial}{\partial \mathbf{x}_i}\left(\rho{(\gamma\cdot D}_m+D_t)\frac{\partial C}{\partial \mathbf{x}_i}\right). \label{eq:3}
\end{equation}
Eqs.~\eqref{eq:1}--\eqref{eq:3} represent the continuity equation, momentum conservation equation, and species conservation equation, respectively. Here, \( p \), \( u \), and \(\mathbf{x} \) denote pressure, velocity, and coordinate position, respectively. \( C \) represents the time-averaged concentration. \( \rho \), \( \mu_m \), and \( D_m \) are physical constants representing density, laminar viscosity, and laminar mass diffusivity, respectively. \( \mu_t \) and \( D_t \) represent turbulent viscosity and turbulent mass diffusivity, which can be obtained from the $k$-$\varepsilon$ two-equation model~\cite{launder1983numerical} and $\overline{c^{\prime2}}$-$\varepsilon_{c^\prime}$ two-equation model~\cite{yu2014introduction}, respectively. The two-equation models and their parameters can be found in Ref.~\cite{jia2022renormalization}. Eqs.~\eqref{eq:1}--\eqref{eq:2} and the closure equation for turbulent viscosity \( \mu_t \) constitute the CFD equations used to calculate the velocity and pressure distributions. The CMT equations include the CFD equations, Eq.~\eqref{eq:3} for concentration calculation, along with the closure equations for turbulent mass diffusivity \( D_t \).

Additionally, to account for the influence of changing the system's topological structure on the physical fields, we introduce a source term \( f_i \) in the momentum conservation equation (Eq.~\eqref{eq:2}) to represent the resistance effect of solid baffles on the flow field. Moreover, in the component conservation equation, a fluid volume fraction \( \gamma \) within the range [0, 1] is introduced to represent the effect of the solid baffle on the component diffusion coefficient. Specifically, \( f_i \) represents the frictional drag generated by solid baffles and can be solved using a density model~\cite{jia2021multi}:
\begin{equation*}
f_i=-\alpha u_i, \label{eq:4}
\end{equation*}
where \( \alpha \) is the interpolation function of the porosity \( \gamma \):
\begin{equation*}
\alpha=\alpha_{\text{min}}+(\alpha_{\text{max}}-\alpha_{\text{min}})q\frac{1-\gamma}{1+\gamma}. \label{eq:5}
\end{equation*}
\( \alpha \) represents the solid density, and \( q \) is a positive integer used to adjust the shape of the interpolation curve. In this study, \( q \) and \( \alpha_{\text{min}} \) are set to 1 and 0, respectively~\cite{jia2022investigation}. \( \alpha_{\text{max}} \) is a large value that approximates the fluid-to-solid transition, set to 600,000. When \( \gamma \) is 0, the corresponding fluid density is \( \alpha_{\text{max}} \), resulting in significant frictional resistance \( f_i \), making the domain effectively solid-like. Conversely, when \( \gamma \) is 1, the corresponding region is fluid. In this study, we set the velocity inlet and pressure outlet (with static pressure fixed at 0)~\cite{kou2024physics}. More computational details can be found in Ref.~\cite{kou2024physics}.
 
\subsection{Data generation}
\label{sec:data-generation}

To train the neural operators, various topological structures $\gamma(x,y)$ are randomly generated and evenly assigned to 9 different inlet velocities (\( v \in \{0.1, 0.2, \dots, 0.9\} \, \text{\textit{m/s}} \)). To obtain the corresponding concentration $C$ and pressure $p$ distributions for the given \( v \) and \(\gamma(x, y)\), the mechanism model (Section~\ref{sec:mechanism-model}) is solved in the
commercial CFD software package FLUENT
\(14.5^{TM}\). The value of \(\gamma(x, y)\) is stored using user-defined memory, while \( \mu_t \), \( D_t \), and \( f_i \) in Eqs.~\eqref{eq:2}--\eqref{eq:3} are implemented through user-defined functions. The total pressure \( p \), including static and dynamic pressures, is normalized as \( P = \frac{p}{2000} + 0.26 \). Each data point in the dataset includes the coordinates \((x, y)\), along with their corresponding values of \((\gamma, P, C)\). The initial dataset is categorized into training, testing, and interpolation/extrapolation datasets based on the value of the inlet velocity \( v \). More details on the initial dataset are provided in Section~\ref{ap:Initial dataset}.

\subsection{Neural operator}
\label{sec:neural operator}

Fourier-DeepONet, illustrated in Fig.~\ref{fig:main2} Step 1, was developed based on three neural network frameworks: the deep operator network (DeepONet)~\cite{lu2021learning}, the Fourier neural operator (FNO)~\cite{li2020fourier}, and U-FNO (a block combining Fourier neural operator and U-Net)~\cite{wen2022u}. DeepONet is used to encode the input variables and functions, while FNO and U-FNO apply Fourier transforms to decode DeepONet's output to obtain the model output. Fourier-DeepONet exhibits superior generalizability and better accuracy in learning neural operators in high-dimensional spaces~\cite{zhu2023fourier}.

The vanilla DeepONet consists of two components: a branch net and a trunk net used for encoding the inputs. In this study, the trunk net takes the inlet velocity \( v \) as input, while the branch net takes the distribution of fluid volume \( \gamma(x,y) \) of the design domain \( \Omega \) (Fig.~\ref{fig:main1}a) as input. Since the solutions of the mechanism model are used as data for training Fourier-DeepONet, the dimension of the input function \( \gamma(x,y) \) corresponds to the number of grid nodes of the mechanism model, which is 121 (mesh nodes in the $x$-direction) \(\times\) 101 (mesh nodes in the $y$-direction). The outputs of the branch net and the trunk net are denoted as \( \mathbf{b} \) and \( \mathbf{t} \), respectively:
\begin{equation*}
\mathbf{b}=B[P\left \langle \gamma(x,y) \right \rangle]\in \mathbb{R}^{L_0\times H_0\times C}, \label{eq:13}
\end{equation*}
\begin{equation*}
\mathbf{t}=T(v)\in\mathbb{R}^C, \label{eq:14}
\end{equation*}
where \( B \) and \( T \) represent two linear transformations, and \( P\left \langle \cdot \right \rangle \) denotes the padding operation. \( L_0 \times H_0 \) represents the output dimensionality of a single channel in the branch net, where in this study \( L_0 = 128 \) and \( H_0 = 112 \). \( C \) represents the number of channels, which corresponds to the width of the operator layers. Additionally, as shown in Fig.~\ref{fig:main2}, DeepONet computes \( \mathbf{z}_0 \) by combining the outputs \( \mathbf{b} \) and \( \mathbf{t} \) through element-wise multiplication:  
\begin{equation*}  
\mathbf{z}_0 = \mathbf{b} \odot \mathbf{t}. \label{eq:15}  
\end{equation*}
 
We then utilize an FNO layer and a U-FNO layer to decode the output of DeepONet. The outputs of the FNO layer, \( \mathbf{z}_1 \), and the two U-FNO layers, \( \mathbf{z}_2 \) and \( \mathbf{z}_3 \), are computed as
\begin{equation*}
\mathbf{z}_1=\sigma(\mathcal{F}^{-1}\left(R_1\cdot\mathcal{F}\left(\mathbf{z}_0\right)\right)+W_1\mathbf{z}_0+b_1), \label{eq:16}
\end{equation*}
\begin{equation*}
\mathbf{z}_2=\sigma(W_2^\prime(\mathcal{F}^{-1}\left(R_2\cdot\mathcal{F}\left(\mathbf{z}_1\right)\right)+\mathcal{U}_2\left(\mathbf{z}_1\right)+W_2\mathbf{z}_1+b_2)), \label{eq:17}
\end{equation*}
\begin{equation*}
\mathbf{z}_3=\sigma(W_3^\prime(\mathcal{F}^{-1}\left(R_3\cdot\mathcal{F}\left(\mathbf{z}_2\right)\right)+\mathcal{U}_3\left(\mathbf{z}_2\right)+W_3\mathbf{z}_2+b_3)). \label{eq:18}
\end{equation*}
In the FNO and U-FNO layers, \( \mathcal{F} \) represents the two-dimensional Fast Fourier Transform (FFT), \( \mathcal{F}^{-1} \) denotes the inverse two-dimensional FFT, \( \sigma \) is the activation function, \( W_i \) are weight matrices, \( R_i \) are complex valued tensors, and \( b_i \) are bias. In the U-FNO layer, \( \mathcal{U}_i \) denotes a U-Net layer. Since the input and output dimensions of the Fourier and U-Net layers must be consistent, a linear layer is added before the activation function in the U-FNO layer to transform the output function to the same dimension as the concentration and pressure distributions. \( W'_{i} \) represents the weight matrix of this linear layer.
 
The projection layer in the decoding part is used to perform nonlinear transformations and slicing operations on the decoded output function \( \mathbf{z}_3 \) from the Fourier layer, resulting in the model's output, i.e., the physical fields \( U(x,y) = C\) or $P$. The projection layer can be expressed as
\begin{equation*}
U\left(x,y\right)=S\left \langle (W_{P2}\sigma(W_{P1}\mathbf{z}_3+b_{P1})+b_{P2}) \right \rangle, \label{eq:19}
\end{equation*}
where $S\left \langle \cdot \right \rangle$ denotes the slicing operation, \( W_{P1} \) and \( W_{P2} \) are the weight matrices, and \( b_{P1} \) and \( b_{P2} \) are biases of the projection layer. The dimensions of the concentration distribution \( C(x,y) \) and the pressure distribution \( P(x,y) \) are consistent with the number of nodes in the CFD simulation, i.e., 201 (number of nodes in the $x$-direction) \(\times\) 101 (number of nodes in the $y$-direction).

As shown in Fig.~\ref{fig:main1}c, the pressure distribution is more sensitive to changes in the topological structure than the concentration distribution, which requires more channels for \( \mathcal{G}_P \) than for \( \mathcal{G}_C \). In this study, the number of channels used for \( \mathcal{G}_C \) and \( \mathcal{G}_P \) was set to 32 and 48, respectively. For other model hyperparameters of DeepONet and the projection layer \( Q \), please refer to Refs.~\cite{lu2021learning,zhu2023fourier}, and for other model hyperparameters of FNO and U-FNO, refer to Refs.~\cite{wen2022u, li2020fourier}. The training loss trajectories of \( \mathcal{G}_C \) and \( \mathcal{G}_P \) are shown in Section~\ref{ap:algorithm figures} Fig.~\ref{fig:A5}.

\subsection{Optimization of neural topologies}
\label{sec:optimization algorithm}

As shown in Fig.~\ref{fig:main2} Step 2, a topological structure is constructed by a neural network, which consists of two layers, each containing 140 neurons. \( \gamma_0 \) represents the preliminary predicted porosity values by the neural network. As the porosity \( \gamma \) must take values of 0 or 1 for any input \( (v, x, y) \), to obtain a reasonable topological structure, the preliminary output \( \gamma_0 \) needs to be transformed to 0/1. In this study, the sigmoid function is used for the 0/1 transformation, and the topological structure is computed as
\begin{equation*}
 \gamma(x,y)=sigmoid(\alpha\cdot(\gamma_0(x,y)-\bar{\gamma_0})), \label{eq:22}
\end{equation*}
where \( \bar{\gamma_0} \) represents the average value of \( \gamma_0 \) at different coordinate points \( (x, y) \)  for the same velocity boundary. \( \alpha \) is a hyperparameter, and a larger value of \( \alpha \) indicates a sharper change between the output of 0 and 1. In this study, \( \alpha = 10 \).

The trained \( \mathcal{G}_P \) and \( \mathcal{G}_C \) neural operators take the \( v \) and the topological structure as the inputs to predict the pressure and concentration distributions, respectively. Then \( \sigma_C^2 \), \( \Delta P \), and \( Obj \) are computed according to Eqs.~\eqref{eq:7}--\eqref{eq:9}. The training losses \(\mathcal{L}_{\sigma_C^2}\), \(\mathcal{L}_{\Delta P}\), and \(\mathcal{L}_{obj}\) are computed from \( \sigma_C^2 \), \( \Delta P \), and \( Obj \) under different inlet velocities:
\begin{equation*}
 \mathcal{L}_{\sigma_C^2}=\frac{1}{N}\sum_{n=1}^{N}{{\sigma_C^2}^{(n)}},
\label{eq:30}
\end{equation*}
 \begin{equation*}
 \mathcal{L}_{\Delta P}=\frac{1}{N}\sum_{n=1}^{N}{{\Delta P}^{(n)}},
\label{eq:31}
\end{equation*}
 \begin{equation*}
 \mathcal{L}_{obj}=\frac{1}{N}\sum_{n=1}^{N}{Obj^{(n)}},
\label{eq:24}
\end{equation*}
where the superscript \( n \) represents different data points in the training dataset. The total inequality constraint of the solid volume \( \mathcal{L}_{vol} \) is computed based on Eq.~\eqref{eq:10}:
\begin{equation*}
 \mathcal{L}_{vol} = \frac{1}{N} \sum_{n=1}^{N} \min \left\{ \iint_\Omega \left( 1 - \gamma^{(n)} \right) dV - 0.1 \times V_\Omega, 0 \right\}.
\label{eq:25}
\end{equation*}
The optimization of a neural topology does not require observational data, and the training loss \( \mathcal{L} \) is composed of topological structure constraints and the objective function:
\begin{equation*}
\mathcal{L} = \mathcal{L}_{obj}+\lambda\cdot \mathcal{L}_{vol}, \label{eq:23}
\end{equation*}
where \( \lambda \) is a weight coefficient and is set to \(1 \times 10^{3}\) to ensure the optimal topology satisfies the volume constraint.

During the training, the initial learning rate decays from \(1 \times 10^{-3}\) to \(1 \times 10^{-4}\) in 5000 steps. The activation function and optimizer are ReLU and Adam, respectively. When \( \omega = 1 \), the training error curve of the TO framework is shown in Fig.~\ref{fig:A7}. We observe that \( \mathcal{L}_{vol} \) approaches 0, indicating that the TO results satisfy the inequality constraints of solid volume.

\subsection{Active data augmentation}
\label{sec:active learning}

After performing Step 1 and Step 2 using the initial dataset in Section~\ref{ap:Initial dataset}, the TO result I is obtained, and some optimized structures are depicted in Fig.~\ref{fig:A3}a. In Step 2, we use neural operators to replace the mechanism model for enabling fast prediction of the objectives. Hence, a good prediction accuracy of neural operators is crucial, and the errors are listed in Table~\ref{tab:A2}. \( \mathcal{G}_P \) shows lower prediction accuracy for the optimized topological structure compared to \( \mathcal{G}_C \), because the training dataset size of \( \mathcal{G}_P \) is equal to the number of topologies in the dataset, regardless of the inlet velocities, and thus it is much smaller than the dataset size of \( \mathcal{G}_C \). Moreover, as suggested by the TO result I, the optimal structure would only have one or two solid structures near the bottom or left boundaries of the design domain. However, in our randomly-generated initial dataset, the structures do not satisfy this pattern.

To improve the prediction accuracy of \( \mathcal{G}_P \), especially for structures similar to the optimal structure, more data are generated as follows. In the first round of data augmentation, the solid positions are chosen the same as those in Fig.~\ref{fig:A3}a, while the solid shapes are randomly generated. We generate ten new structures for \( v = 0.1, 0.2, \dots, 0.9 \, \mathrm{m/s} \), and we show three examples in Fig.~\ref{fig:A4}. To reduce the computational cost, we only use these structures to generate the training data of \( \mathcal{G}_P \). Then, TO result II is obtained through transfer learning of the \( \mathcal{G}_P \) neural operator and the neural topology.

In the second round of data augmentation, we use the 27 optimal structures obtained from TO result II to generate new training data for both \( \mathcal{G}_C \) and \( \mathcal{G}_P \).

\section*{Data availability}
The simulation and experiment data will be made available upon publication on GitHub at \url{https://github.com/lu-group/neural-topology-optimization}. 

\section*{Code availability}
The code for this study is implemented using the Python library DeepXDE~\cite{lu2021deepxde} and will be publicly available at the GitHub repository \url{https://github.com/lu-group/neural-topology-optimization}. 

\section*{Acknowledgments}
This work was partially supported by the NNSFC Grants No.~22178247 (to X.Y.) and No.~22308251 (to S.J.). 

\bibliographystyle{unsrt}
\bibliography{main}

\appendix
\clearpage
\renewcommand{\thesection}{S\arabic{section}}
\renewcommand{\thefigure}{S\arabic{figure}}
\renewcommand{\thetable}{S\arabic{table}}
\setcounter{figure}{0}
\setcounter{table}{0}

\renewcommand*{\theHfigure}{\thefigure}
\renewcommand*{\theHtable}{\thetable}

\section{Initial dataset}
\label{ap:Initial dataset}

\paragraph{Training dataset.}
Since the pressure distribution is more sensitive to variations in the topology structure than the concentration distribution, we use a larger dataset to train the \( \mathcal{G}_P \) neural operator than $\mathcal{G}_C$. Specifically, we randomly generate 891 topological structures, every 89 structures corresponding to one value of inlet velocity \( v \in \{0.1, 0.2, \dots, 0.9\} \,\mathrm{m/s} \). The CFD equations are used to solve the pressure distributions, yielding 891 data for training the \( \mathcal{G}_P \) neural operator. Then, 351 cases, every 39 structures corresponding to one value of inlet velocity \( v \in \{0.1, 0.2, \dots, 0.9\} \,\mathrm{m/s} \), are randomly selected to solve the CMT equations and obtain concentration distributions, forming the training dataset for the \( \mathcal{G}_C \) neural operator. The 9 cases of the smooth channel (\( \gamma(x,y) \equiv 1 \)) with 9 inlet velocities are also included in the training dataset for both \( \mathcal{G}_P \) and \( \mathcal{G}_C \). In total, the training dataset for the \( \mathcal{G}_P \) neural operator consists of 900 cases, and the training set for the \( \mathcal{G}_C \) neural operator consists of 360 cases.

\paragraph{Test dataset.}
We randomly generate 18 topological structures, every two structures corresponding to one value of \( v \in \{0.1, 0.2, \dots, 0.9\} \,\mathrm{m/s} \). Numerical methods were employed to solve the CMT equation system and obtain the corresponding pressure and concentration distributions. These datasets serve as the test datasets for the \( \mathcal{G}_P \) and \( \mathcal{G}_C \) neural operators.

\paragraph{Interpolation dataset.}
To evaluate the predictive accuracy of the neural operators for pressure and concentration distributions at \( v \notin \{0.1, 0.2, \dots, 0.9\}\,\mathrm{m/s} \), we randomly generate 16 topological structures, every two structures corresponding to one value of \( v \in \{0.15, 0.25, \dots, 0.85\}\,\mathrm{m/s} \). This interpolation dataset tests the neural operators' ability to interpolate velocity boundary conditions within the training range.

\paragraph{Extrapolation dataset.} To evaluate the neural operators' ability to extrapolate velocity boundary conditions outside the training range at \( v \notin [0.1, 0.9]\,\mathrm{m/s} \), we randomly generate 4 topological structures, every two structures corresponding to one value of \( v \in \{0.05, 0.95\}\,\mathrm{m/s} \).
 
\clearpage
\section{Validation of neural operators}
\label{ap:NO validation}

\begin{figure}[htbp]
    \centering
    \includegraphics[width=1\linewidth]{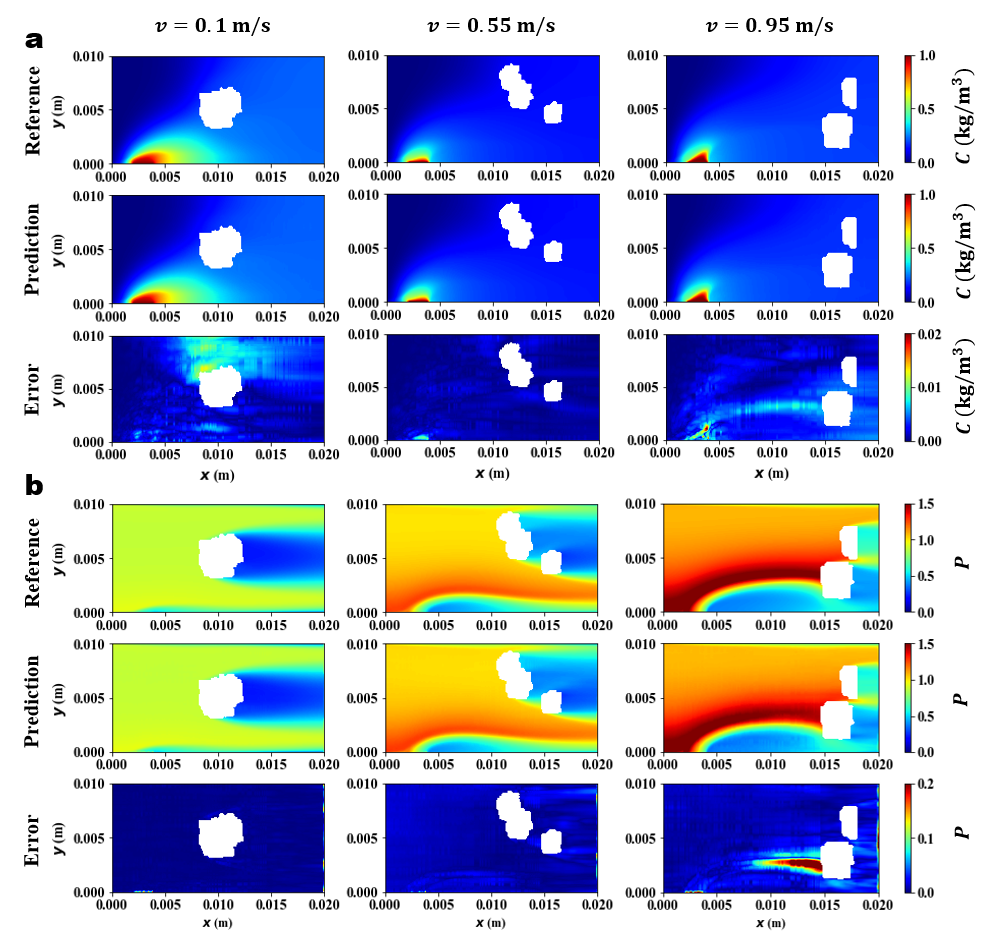}
    \caption{\textbf{Validation of the neural operator predictions.} \textbf{a}, $\mathcal{G}_C$ results. \textbf{b}, $\mathcal{G}_P$ results. Test dataset example with $v= 0.1\,\mathrm{m/s}$; interpolation dataset example with $v= 0.55\,\mathrm{m/s}$; and extrapolation dataset example with $v= 0.95\,\mathrm{m/s}$.}
    \label{fig:A2}
\end{figure}

To obtain the pre-trained neural operators in the TO framework, the \( \mathcal{G}_P \) and \( \mathcal{G}_C \) neural operators were trained using the initial training dataset. The test, interpolation, and extrapolation datasets were then used to validate the generalization ability of the neural operators. Here, we visualize the ground truth, \( \mathcal{G}_P \)/\( \mathcal{G}_C \) network prediction, and error for one case from each of the three datasets (Fig.~\ref{fig:A2}). The predictions of the neural operators are in good agreement with the reference solutions. This demonstrates that the neural operator achieves high prediction accuracy across different topological structures and also exhibits predictive capability for cases with interpolated or extrapolated inlet velocities.

\clearpage
\section{Training procedure}
\label{ap:algorithm figures}

\begin{figure}[htbp]
    \centering
    \includegraphics[width=0.8\linewidth]{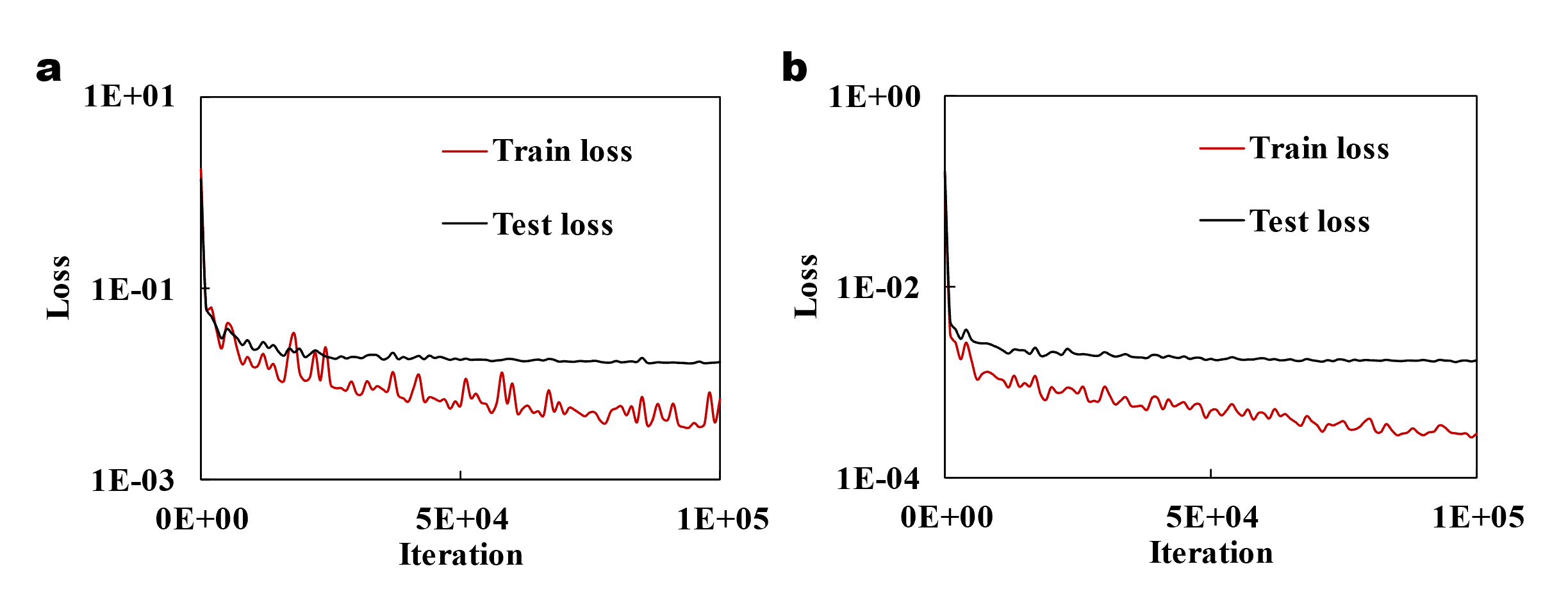}
    \caption{\textbf{Training and test losses for neural operators using the initial training dataset.} \textbf{a}, \( \mathcal{G}_P \) neural operator. \textbf{b}, \( \mathcal{G}_C \) neural operator.}
    \label{fig:A5}
\end{figure}

\begin{figure}[htbp]
    \centering
    \includegraphics[width=0.4\linewidth]{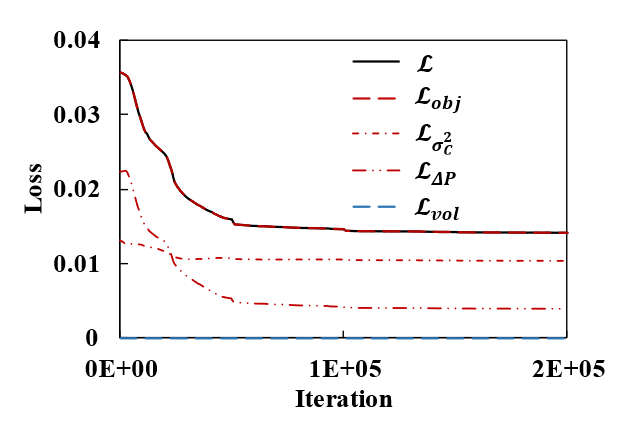}
        \caption{\textbf{Training losses of neural topology optimization with $\omega=1$.}}
    \label{fig:A7}
\end{figure}

Here, we show the losses during network training, including the losses of the neural operators (Fig.~\ref{fig:A5}) and the neural topology optimization (Fig.~\ref{fig:A7}).

\clearpage
\section{Comparison of three TO results}
\label{ap:TO results}

\begin{figure}[htbp]
    \centering
    \includegraphics[width=0.8\linewidth]{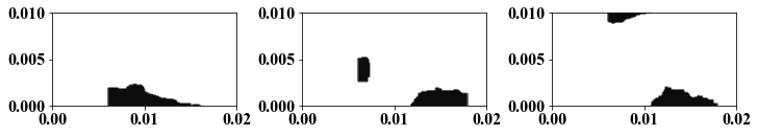}
    \caption{\textbf{Examples of topological structures added in the first round of data augmentation.} The black regions represent the solid baffles, where \(\gamma=1\).}
    \label{fig:A4}
\end{figure}

\begin{figure}[htbp]
    \centering
    \includegraphics[width=0.8\linewidth]{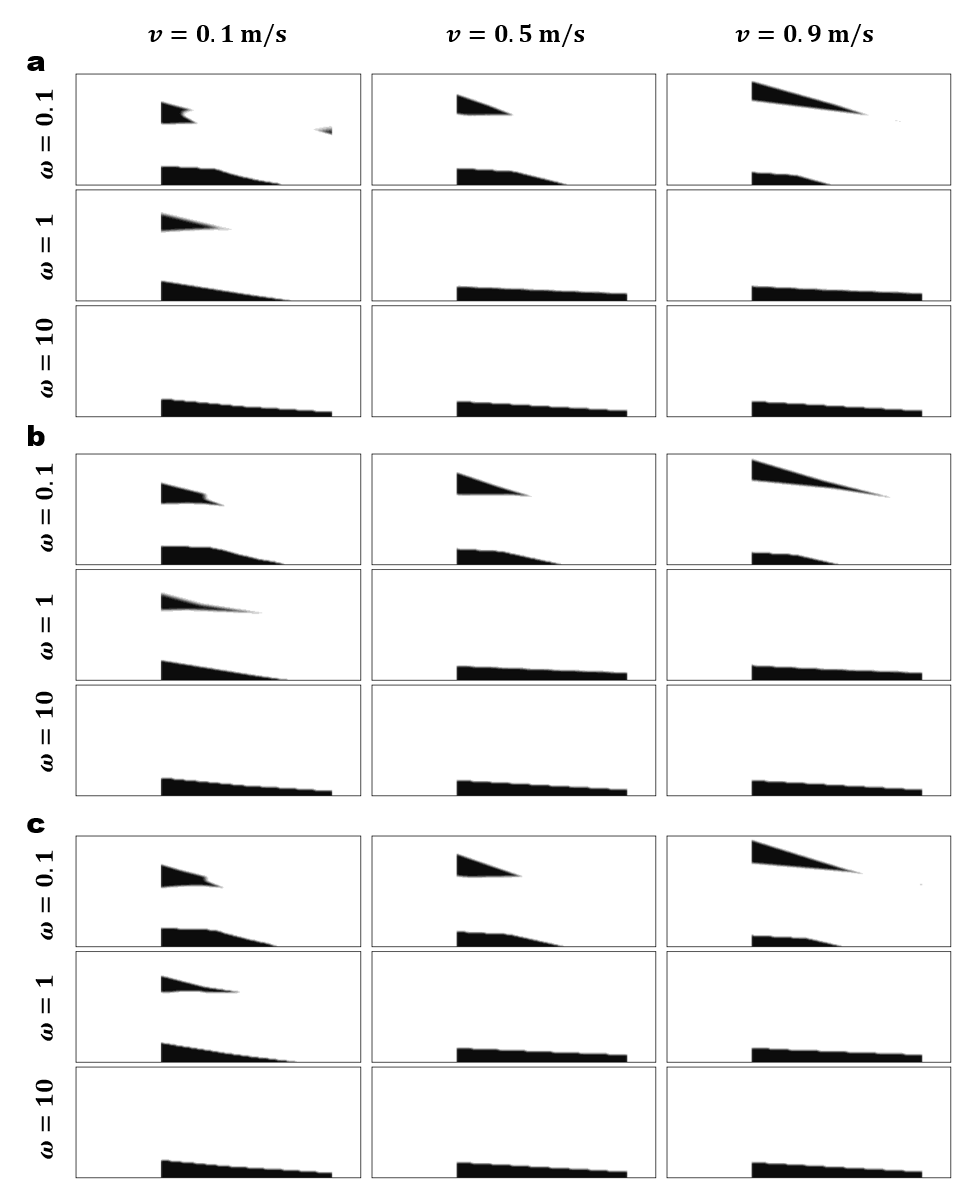} 
    \caption{\textbf{Optimal topological structures for different inlet velocities and $\omega$.} \textbf{a}, TO result I. \textbf{b}, TO result II. \textbf{c}, Final TO result. The black region represents the solid area, where \(\gamma=1\).}  
    \label{fig:A3}
\end{figure}

To validate the effectiveness of the TO algorithm, we use the mechanism model to verify the accuracy of the neural operator’s predictions for the optimal topological structures (Table~\ref{tab:A2}). In the active learning-based TO approach, the optimal topological structure is iteratively improved, and we show three topology structure examples of the first active data augmentation in Fig.~\ref{fig:A4}. We also illustrate the evolution of the optimal topological structures under different weights and inlet velocities in Fig.~\ref{fig:A3}.

\begin{table}[htbp]
    \caption{\textbf{Mean relative errors of the physical fields ($C$ or $P$) and objectives (\(\sigma_C^2\) or \(\Delta P\)) for the optimized topological structures under three different inlet velocities.}}
    \label{tab:A2}
    \centering
    \begin{tabularx}{\textwidth}{p{2.5cm} *{7}{>{\centering\arraybackslash}X}}
        \toprule
        & & \multicolumn{3}{c}{Physical field error} &\multicolumn{3}{c}{Objective error} \\
        \cmidrule(lr){3-5} \cmidrule(lr){6-8} 
        \multicolumn{2}{l}{Inlet velocity \(v\,\mathrm{(m/s)}\)}& 0.1 & 0.5 & 0.9 & 0.1 & 0.5 & 0.9 \\
        \midrule
        \multirow{2}{*}{TO result I}
        &\(C/\sigma_C^2\) & 3.281\%  & 2.504\% & 2.721\% & 0.547\% & 0.372\% & 2.119\%\\
        &\(P/\Delta P\)& 7.747\%  & 5.254\% & 3.855\% & 16.32\% & 7.058\% & 3.577\%\\
        \midrule
        \multirow{2}{*}{TO result II}
        &\(C/\sigma_C^2\)&3.728\%  & 2.506\% & 2.746\% & 0.470\% & 0.350\% & 2.520\%\\
        &\(P/\Delta P\)&6.040\%  & 2.613\% & 1.943\% & 7.286\% & 1.894\% & 2.560\%\\
        \midrule
        \multirow{2}{*}{Final TO result}
        &\(C/\sigma_C^2\) &0.515\%  & 0.294\% & 0.160\% & 0.259\% & 0.253\% & 0.257\%\\
        &\(P/\Delta P\)&0.537\%  & 0.346\% & 0.546\% & 0.550\% & 0.335\% & 0.480\%\\
        \bottomrule
    \end{tabularx}
\end{table}

\clearpage
\section{Experimental data}
\label{ap:experiment}

\begin{figure}[htbp]
    \centering
    \includegraphics[width=1\linewidth]{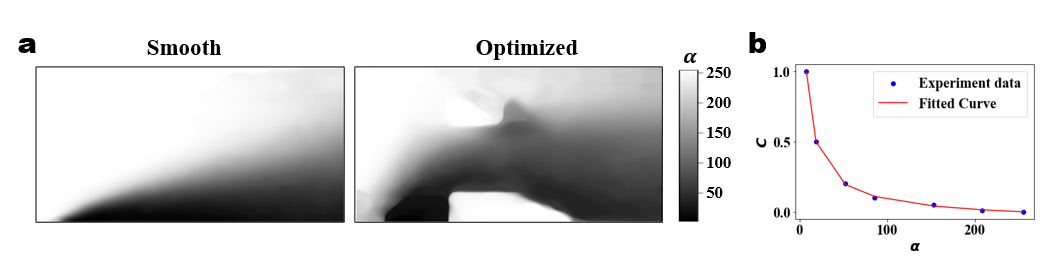}
    \caption{\textbf{Experimental data.} \textbf{a}, Time-averaged gray images of the smooth channel and optimized channel. \textbf{b}, Gray ($\alpha$)--concentration ($C$) calibration curve.}
    \label{fig:A6}
\end{figure}

The experimental measurement data for both the smooth and optimized channel structures can be accessed at
\href{https://drive.google.com/drive/folders/1bkplNOv3o2nkb6ZDkVZyypFc_b2VUdm0?usp=sharing}{\textbf{experiment data}}. The corresponding time-averaged gray images of the smooth and optimized channels are shown in Fig.~\ref{fig:A6}a. We measured the grayscale of standard concentration solutions using the experimental method described in Section~\ref{Experiment setup}, establishing a correlation between grayscale and concentration value (Fig.~\ref{fig:A6}b), which is expressed by the decreasing power function: \(C = 4.7417 \alpha^{-0.7069} - 0.0944
\).

\end{document}